\documentclass[%
 preprint,
showpacs,preprintnumbers,
 amsmath,amssymb,
 aps,
pre,
]{revtex4-1}

\usepackage{amsmath}
\usepackage{amsfonts}
\usepackage{amssymb}
\usepackage[utf8]{inputenc}
\usepackage[english]{babel}
\usepackage{color}
\usepackage{subfigure} 

\usepackage{graphicx}

\begin{document}

\preprint{APS/123-QED}

\title{Wada property in systems with delay}

\author{Alvar~Daza, Alexandre~Wagemakers, Miguel A.F.~Sanju\'an}
 \affiliation{Nonlinear Dynamics, Chaos and Complex Systems Group, Departamento de  F\'isica, Universidad Rey Juan Carlos\\
Tulip\'an s/n, 28933 M\'ostoles, Madrid, Spain\\
alvar.daza@urjc.es, alexandre.wagemakers@urjc.es, miguel.sanjuan@urjc.es}


%
%

\date{\today}

\begin{abstract}
Delay differential equations take into account the transmission time of the information. These delayed signals may turn a predictable system into chaotic, with the usual fractalization of the phase space. In this work, we study the connection between delay and unpredictability, in particular we focus on the Wada property in systems with delay. This topological property gives rise to dramatical changes in the final state for small changes in the history functions. 
\end{abstract}

\maketitle

\section{\label{sec:Introduction}Introduction}

The objective of this work is twofold: on the one hand, we study the effects of the delay in nonlinear dynamical systems, focusing on their associated uncertainty; on the other hand, we study the emergence of the Wada property when delays are involved.

Delay differential equations (DDEs) take into account the time taken by systems to sense and react to the information they receive, in other words, that the information transmission cannot be instantaneous, but delayed. For practical purposes, these lags can often be ignored when their timescales are very small compared to the dynamics of the system. However, there are situations where large delays cannot be overlooked: genetic oscillators \cite{lewis_autoinhibition_2003}, neuron networks \cite{nordenfelt_bursting_2013}, respiratory and hematopoietic diseases \cite{mackey_oscillation_1977}, electronic circuits \cite{wang_generating_2001}, optical devices \cite{ikeda_high-dimensional_1987}, engineering applications \cite{kyrychko_use_2010}, etc. Delay differential equations provide a very useful tool for the modelling of the previous examples. Moreover, they are able to display such interesting kind of dynamics as deterministic brownian motion \cite{sprott_simple_2007}, hyperchaos \cite{ikeda_study_1986} and many cooperative effects \cite{ramana_reddy_experimental_2000, nordenfelt_frequency_2014, lakshmanan_dynamics_2011}. It is also important to mention that DDEs have the property of time-irreversibility \cite{mackey_dynamic_1989}, which introduces further difficulties in the analysis but also gives a more realistic perspective of many processes.

A crucial feature of DDEs is that they need an infinite set of initial conditions to be integrated. This set is usually called {\it history} and provides the state of the system before the action of the delayed terms. Sometimes history is set randomly, although given the sensitivity of some systems with delay and the difficulties arising when an infinite set of initial random points is needed, the choice of random histories is a delicate issue. A better option is to set the history as the solution of the system without the delayed terms. Another possibility are history functions described by some parameters and properly chosen for each physical situation.


A convenient way to handle this infinite number of initial conditions is to define history functions characterized by a finite number of parameters. 
This supposes a huge difference with respect to nonlinear systems modelled with ordinary differential equations (ODEs), since the space of the history functions and the real phase space are not in correspondence. Therefore, basins of attraction, which are a very powerful tool to study sensitivity in dissipative systems, have a different nature in DDEs.

In dissipative dynamical systems defined by ODEs, the basins of attraction register in a plot the attractors reached by different initial conditions. In DDEs the same idea can be exploited: we can plot basins of attraction varying the parameters of the history functions. These basins are subspaces of the infinite dimensional space of history functions, but still can provide much information about the sensitivity of the system. Only a few authors have studied basins of attraction in DDEs \cite{aguirregabiria_fractal_1987, losson_solution_1993, taylor_approximating_2007}. In this work, we study how the delay can induce uncertainty in the system, and we pay special attention to the appearance of the Wada property in these basins of attraction made out of different history functions.

The Wada property is a topological property that can be stated as follows: given more than two open sets, they all share the same boundary. This situation is very counter-intuitive, since most boundaries are only between two sets. In some cases there might be some points or regions that separate more than two open sets, but the case where every point in the boundary is in the boundary of all the sets is unique. 

This topological curiosity was first reported by Kunizo Yoneyama \cite{yoneyama_theory_1917}, who attributed its discovery to his teacher Takeo Wada. Later on, the research work done by James Yorke and coworkers contributed to relate the Wada property with nonlinear dynamics \cite{kennedy_basins_1991, nusse_saddle-node_1995, nusse_wada_1996, nusse_fractal_2000}. They found that the basins of attraction of simple physical systems such as the forced pendulum can possess the Wada property \cite{nusse_wada_1996}. In these seminal works, they gave a topological argument showing that the origin of this property was an unstable manifold crossing all the basins \cite{nusse_saddle-node_1995}. After that, the Wada property has been found in a variety of models associated to physical phenomena \cite{aguirre_unpredictable_2002, aguirre_wada_2001, toroczkai_advection_1998}. 

The interest in the Wada property lies on the fact that their boundaries are the most entangled one can imagine, since they separate all the basins at the same time. Therefore, small perturbations near the boundaries can lead to any of the different attractors of the system. Recently, Zhang and Luo \cite{zhang_wada_2013} have studied an intermediate situation, where some regions have the Wada property but others do not, so they call this situation partial Wada basins. 

The Wada property also appears in systems with more than two degrees of freedom \cite{epureanu_fractal_1998, kovacs_topological_2001, sweet_topology_1999}. In these cases the basins have more than two dimensions and the subspaces generally show the disconnected Wada property: the different basins share the same boundary but they are disconnected. These disconnected Wada basins can be analyzed by means of the techniques developed in \cite{daza_testing_2015}.


The aim of this work is to investigate the connection between delay and unpredictability, looking also for the Wada property in systems with delay. We organize this search as follows. In Sections \ref{sec:LinearDelay} and \ref{sec:NonlinearDelay} we introduce two delayed systems that present different degrees of the Wada property. Finally, in Section \ref{sec:Discussion} we briefly summarize and discuss our main results.

\section{\label{sec:LinearDelay}Forced delayed action oscillator}

\begin{figure}
\begin{center}
\subfigure[]{\includegraphics[width=7cm]{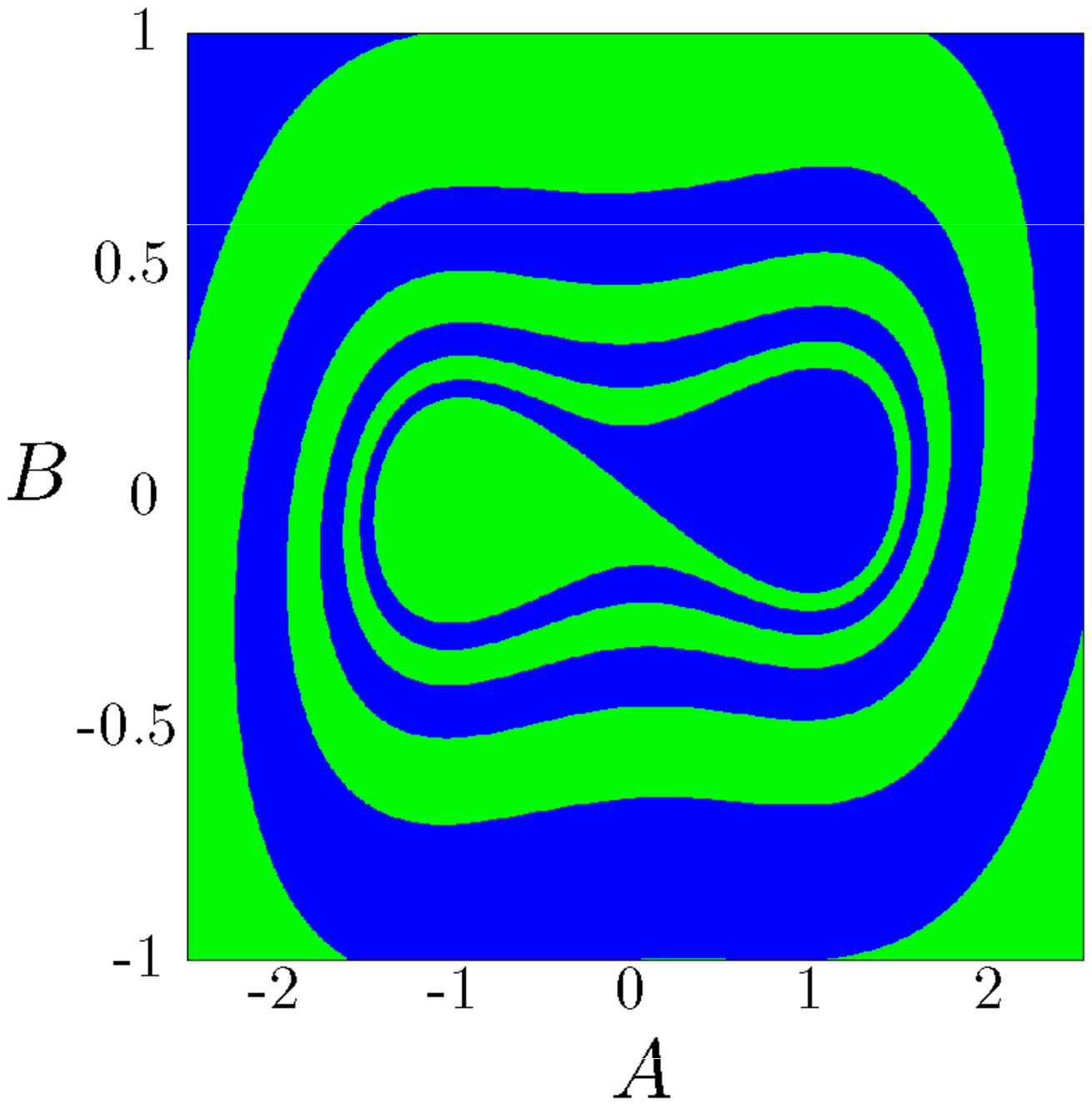}}
\subfigure[]{\includegraphics[width=7cm]{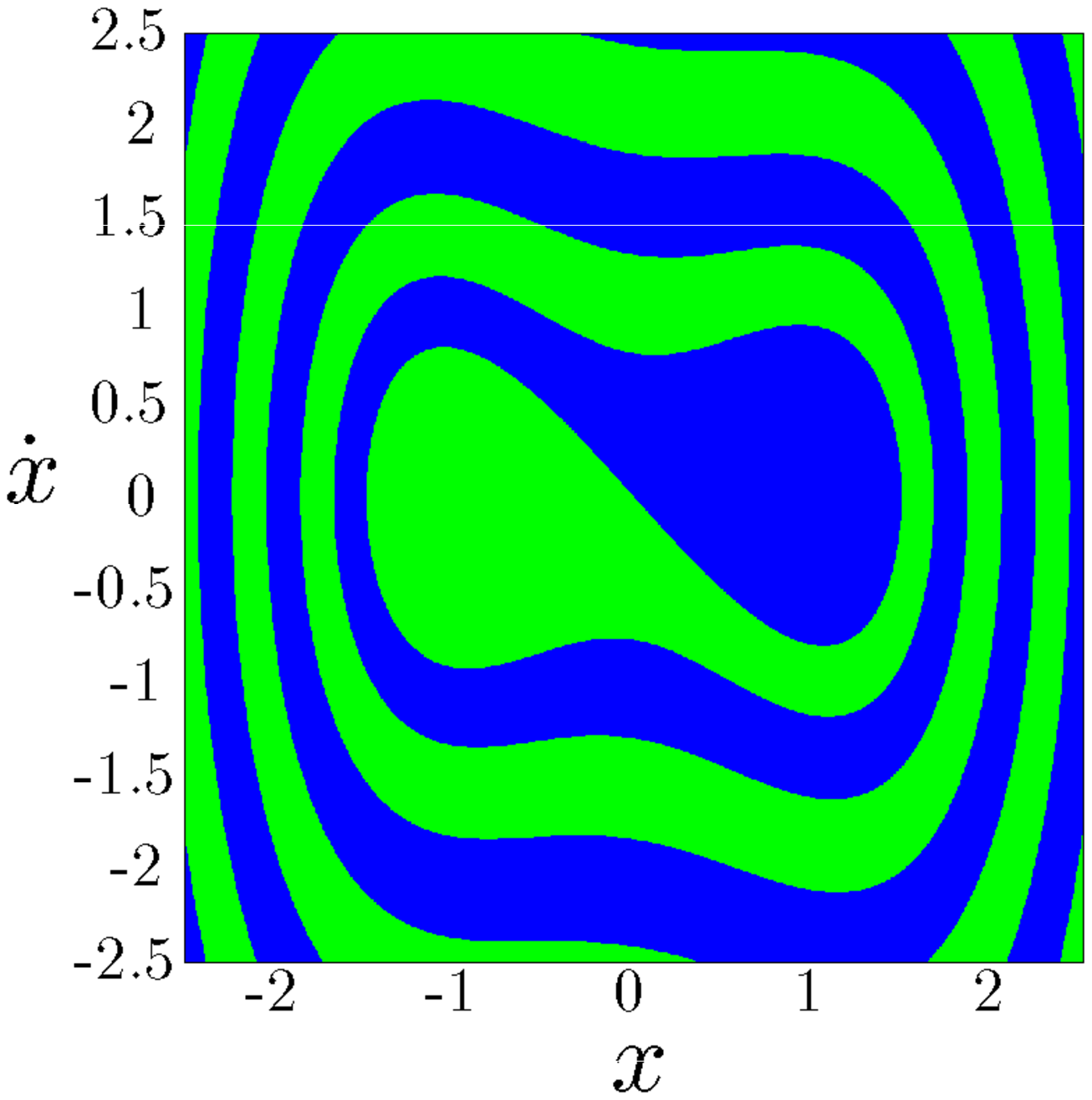}}
\end{center}
\caption{\label{fig:Similar} \textbf{Comparison between the delayed action and Duffing oscillators}. (a) Basin of attraction of the delayed action oscillator $\dot{x}+ x ((1+\alpha) x^2 -1) -\alpha x_{\tau}=0.$ with $\alpha=-0.95$ and $\tau=1.065$. (b) Basin of attraction of the Duffing oscillator $\ddot{x}+\gamma\dot{x}+ x (x^2-1)=0.$ with $\gamma=0.15$. Both figures show a similar topology, but Fig.\ref{fig:Similar} (b) represents only a slice of the infinite dimensional space of history functions, given by the family of history functions of Eq. \ref{eq:History}.}
\end{figure}

We start studying an apparently simple system sometimes called the delayed action oscillator (DAO). It is a single variable system with a double-well potential and a linear delayed feedback with constant time delay $\tau$, that we will denote $x_{\tau}$. It can be stated as follows,

\begin{equation}
\dot{x}+ x ((1+\alpha) x^2 -1) -\alpha x_{\tau}=0. \label{eq:ENSOmodif}
\end{equation}
where $\alpha, \tau \in \mathbb{R}$. Boutle et al. \cite{boutle_nino_2007} proposed this model in the context of the ENSO (El Niño Southern Oscillation) phenomenon, where the variable $x$ represents the temperature anomaly of the ocean's surface. In \cite{redmond_bifurcation_2002} the authors analyze the stability and bifurcations of this system by a center manifold reduction. They demonstrate that as the delay increases beyond a critical value $\tau_c$, the steady state solution $x=0$ can undergo a Hopf bifurcation giving rise to a limit cycle. Without the delayed term, this system would be a one-dimensional ODE and could not oscillate, but the linear delayed feedback makes the system infinite-dimensional allowing oscillatory dynamics. In the case that $\alpha>-1$ and $\tau>\tau_c$ the limit cycle coexists with two stable fixed points, so the system can be multistable. To visualize the situation in a plot, we choose the following family of history functions defined by two parameters $A$ and $B$,
\begin{equation}
x(t)=A+Bt, \: \forall \, t \in \left[ -\tau,t_{0} \right]. \label{eq:History}
\end{equation}
Unless specified, this linear equation will be the family of history functions chosen by default along the paper. Now, we can compute a basin of attraction varying $A$ and $B$. Figure \ref{fig:Similar} (a) represents the basin for $\alpha=-0.95$ and $\tau=1.065$, below the critical value $\tau_c$. It is interesting to notice the analogy between the DAO and the well-known Duffing oscillator
\begin{equation}
\ddot{x}+\gamma\dot{x}+ x (x^2-1)=0. \label{eq:Duffing}
\end{equation}
The structure of its basin of attraction is very similar to the DAO model as shown in Fig. \ref{fig:Similar} (b). However it is important to notice that the two basins have different nature: we have the real phase space for the Duffing oscillator and a slice of the infinite space of history functions in the case of the DAO.

\begin{figure}
\begin{center}
\subfigure[]{\includegraphics[width=6cm]{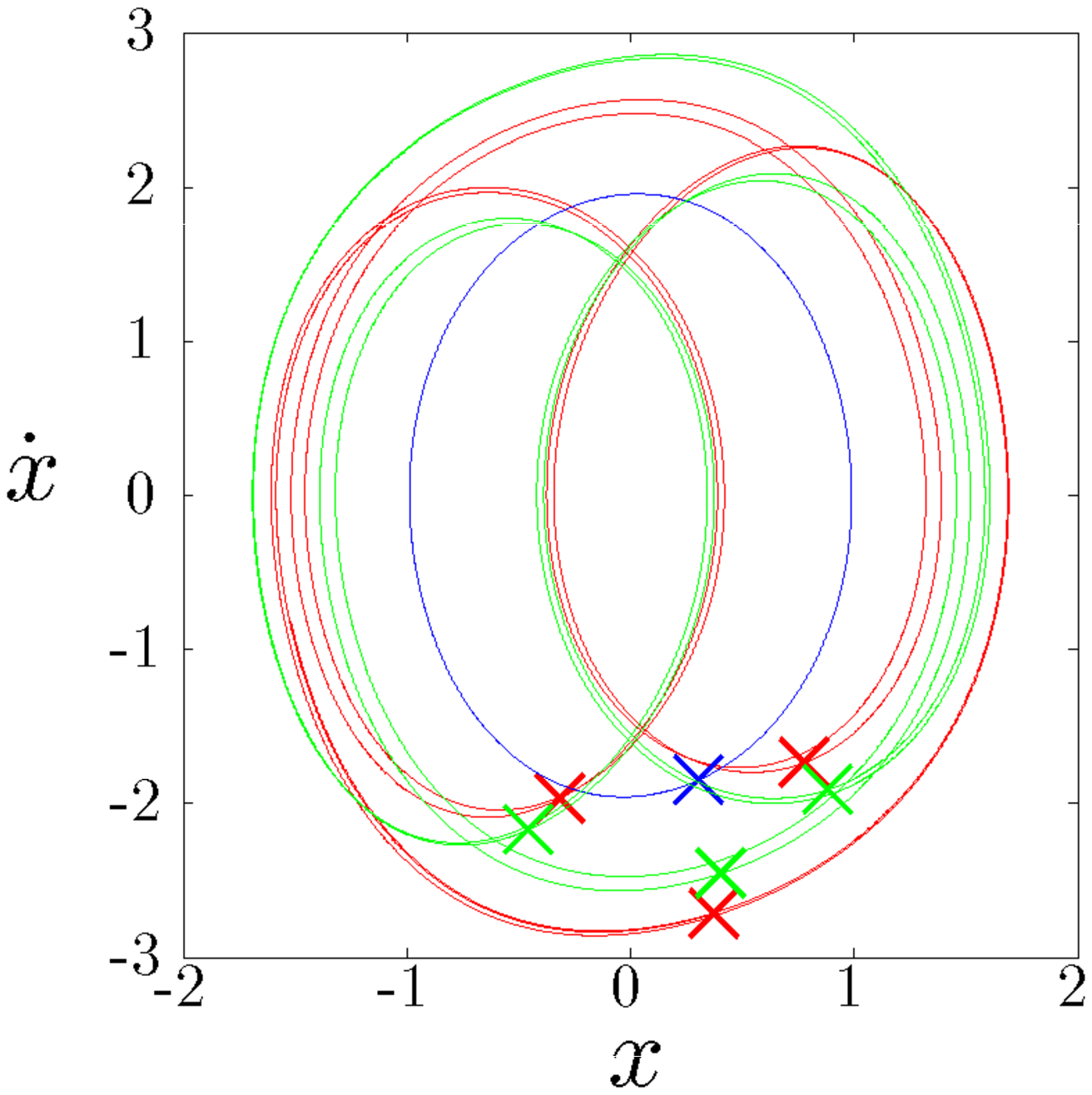}}
\subfigure[]{\includegraphics[width=6cm]{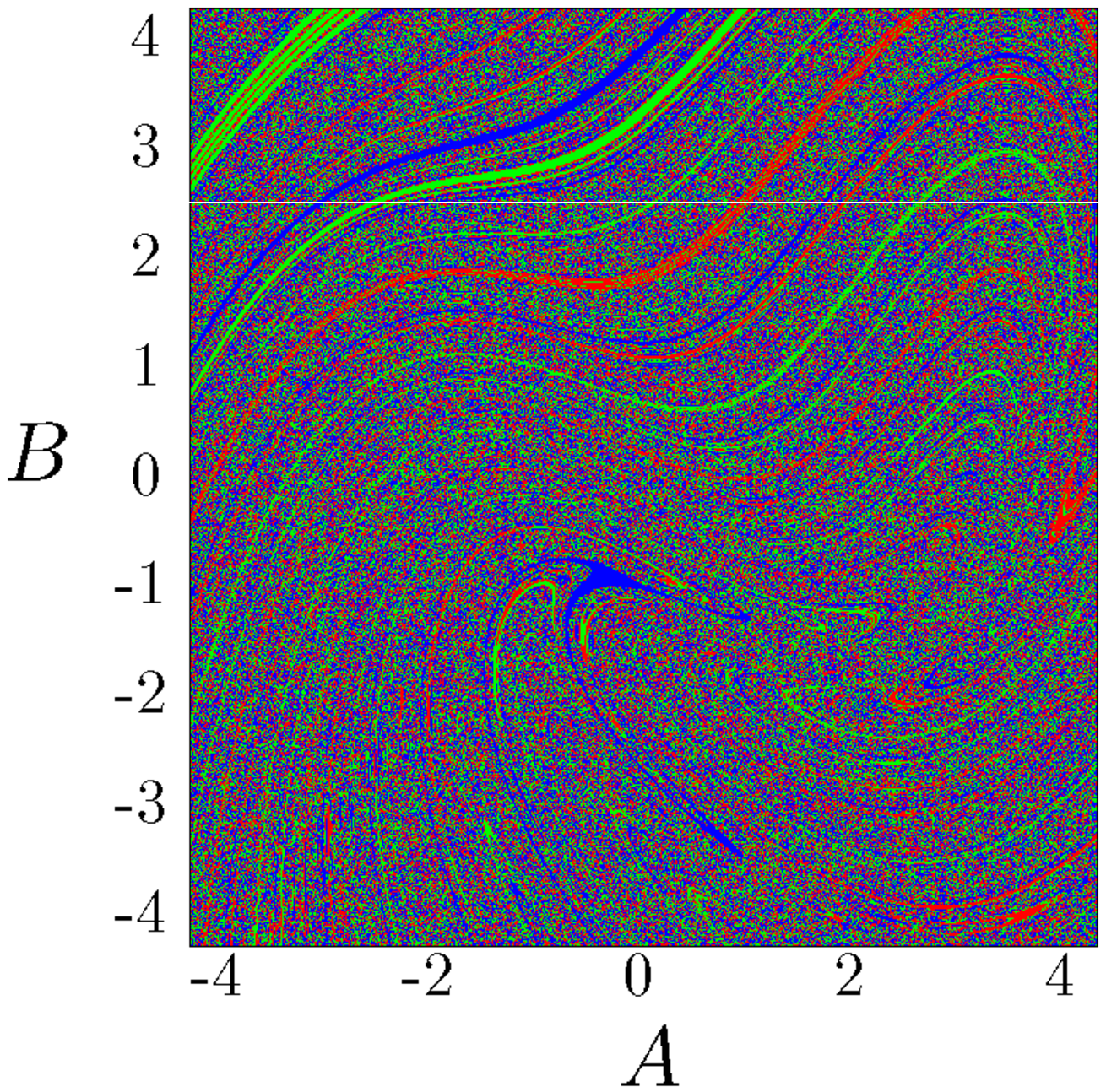}}
\subfigure[]{\includegraphics[width=6cm]{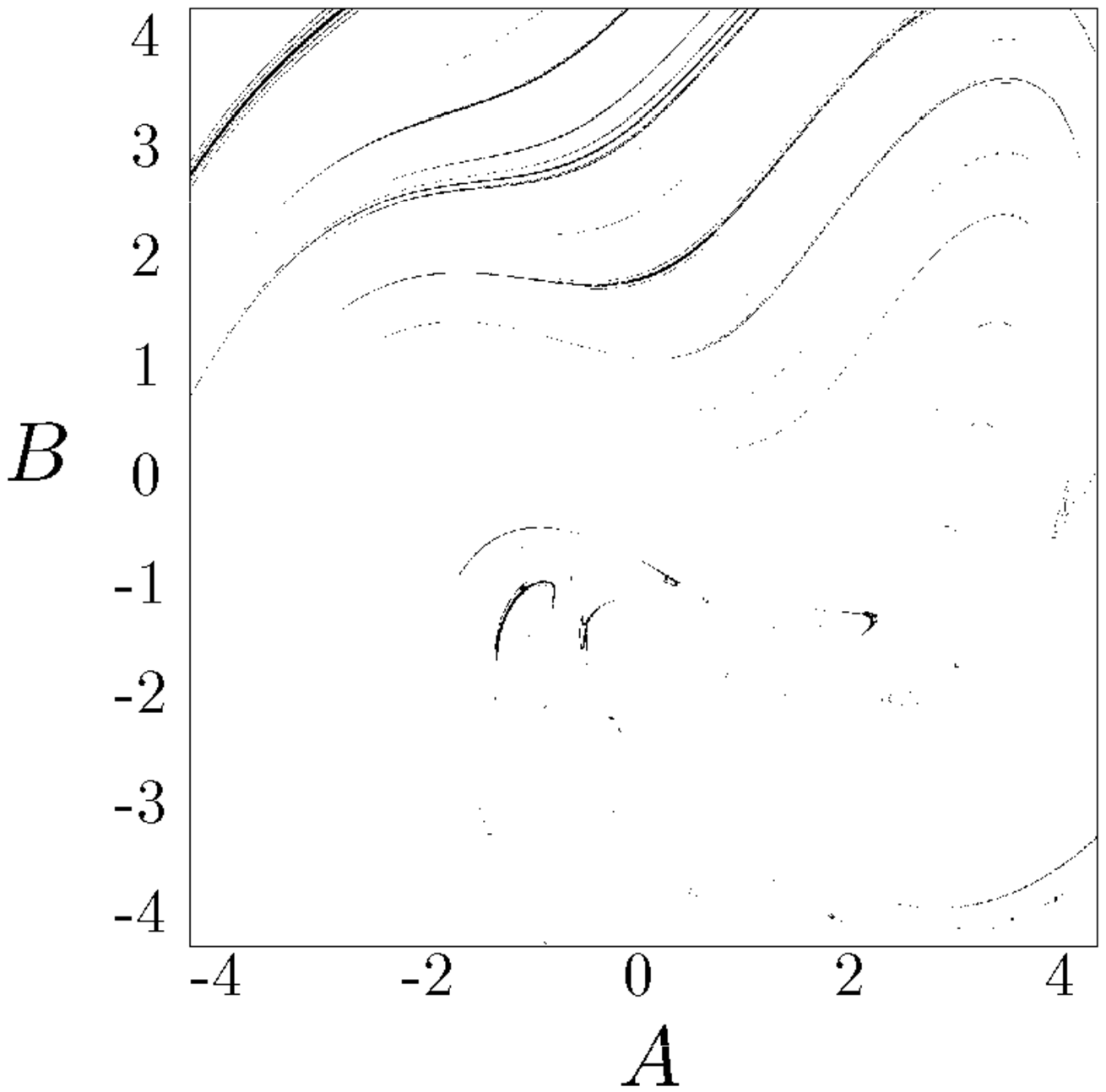}}
\subfigure[]{\includegraphics[width=6cm]{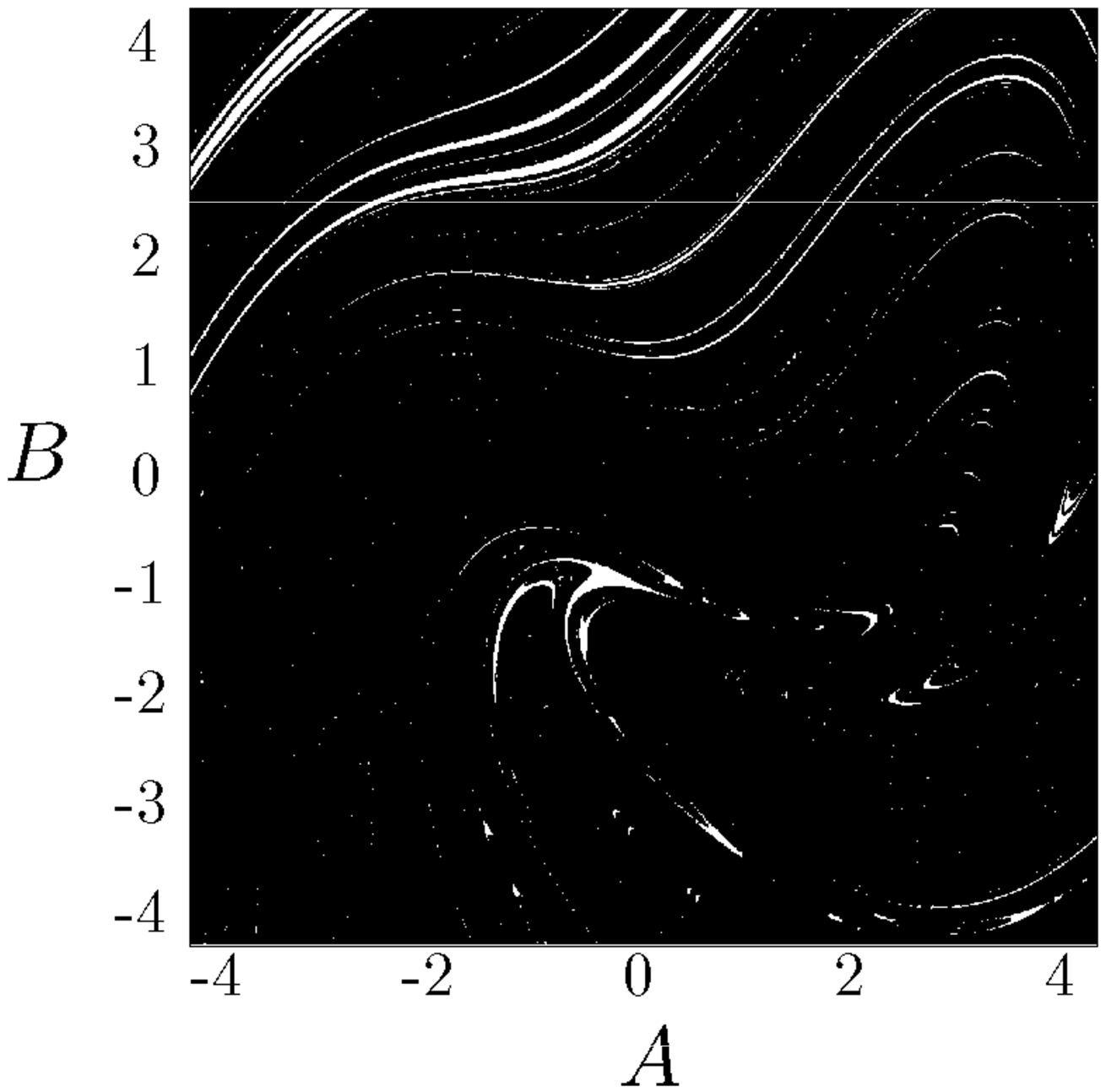}}
\end{center}
\caption{\label{fig:false} \textbf{Transient chaos induced by linear delay}. (a) Attractors of the forced DAO defined by $\dot{x}+ x ((1+\alpha) x^2 -1) -\alpha x_{\tau}=F \sin \omega t$ for $\alpha=-0.925$, $\tau=1.065$, $F=0.525$, and $\omega=1$. There is a period-1 orbit (in the stroboscopic map) and two symmetric period-3 orbits. Trajectories intersect because this is a projection in $(x,\dot{x})$, but the system lives in infinite dimensions in principle. (b) The basins of attraction are highly mixed, increasing the unpredictability of the system. However, as panels (c) and (d) show the system is not completely Wada: some points separate only two basins, so the system is only partially Wada.  }
\end{figure}

At this point it is important to make a connection with Ref. \cite{aguirre_unpredictable_2002}. In that work, Aguirre and Sanju\'an studied the Duffing oscillator driven by a periodic forcing $F \sin \omega t$ on the right hand side of Eq. (\ref{eq:Duffing}). They showed that if the parameters are carefully chosen ($\gamma=0.15, \omega=1, F \in (0.24,0.26)$) the system can display the Wada property. Making a naive analogy, it is plausible that we will encounter the same effect by including the periodic forcing in the DAO such as 
\begin{equation}
\dot{x}+ x ((1+\alpha) x^2 -1) -\alpha x_{\tau}=F \sin \omega t. \label{eq:FalsoWada}
\end{equation}
For $\alpha=-0.925$, $\tau=1.065$, $F=0.525$ and $\omega=1$, this system presents three attractors (see Fig. \ref{fig:false}(a)). Given the periodic forcing, we can make a stroboscopic map taking $t=2\pi n, n \in \mathbb{Z}$. In this map, two of these attractors are period three orbits, suggesting the possibility of chaotic dynamics in the system  \cite{li_period_1975}. In fact, a chaotic attractor exists for other parameters.  For the chosen set of parameters, we can tell that it is the delay that makes possible the appearance of chaotic dynamics in the phase space. Otherwise for $\tau=0$, the system would have a dimension equal to two, forbidding any chaotic motion. 

A few authors have studied the relation between delay and chaos \cite{heiden_dynamics_1982, hale_onset_1988}, but the delay was always considered in a nonlinear term. We show here that a linear delayed feedback can also induce chaos in a continuous system. Some systems displaying chaos have been modified with a linear delayed feedback, preserving the chaotic motion \cite{aguirregabiria_fractal_1987}. But in this model, the linear delayed feedback is providing the extra dimensions that the system needs to display chaos, in the same way that the delay allows sustained oscillations in a system of one variable.

The system of Eq. \ref{eq:FalsoWada} presents transient chaos and possesses three attractors (depicted in Fig. \ref{fig:false}(a)). If we plot the basin of attraction (Fig. \ref{fig:false}(b)) we can see that the picture is highly fractalized and looks like Wada. As discussed before, the Nusse-Yorke method to verify if the basin is actually Wada is not applicable here since we do not have a correspondence between the phase space of the history functions and the actual phase space. However, we can apply our test \cite{daza_testing_2015} in order to decide whether it is Wada or not. 

The algorithm sets a grid and searches for the points lying on the boundary of two or three atractors. The test for Wada is conclusive if all the points in the boundary belong to the boundary of three atractors. Applying our method the 3-boundary has box-counting dimension equal to $1.602$ (see Fig. \ref{fig:false}(c)) and the 2-boundary $0.761$ (see Fig. \ref{fig:false}(d)). The indicator $W_3=0.990$ also reveals that the system is not fully Wada, but only partially Wada. In fact, if we zoom in we can see that the red and green basins do not mix with the blue one, so there is a boundary between red and green that they do not share with the blue basin.

It is clear from the plot of the basin in Fig. \ref{fig:false}(b) that the system is highly unpredictable, but it does not show the Wada property. We have scanned a wide range of parameters and we have not been able to find the Wada property for the forced DAO of Eq. \ref{eq:FalsoWada}. Perhaps there are small parameter ranges where the Wada property arises, but since the requirements that a system must fulfil to exhibit the Wada property are unclear, we cannot assure nor discard that the forced delayed action oscillator can display the Wada property. Nonetheless, the delay can induce not only chaos, but also the Wada property, as we will see in the next section.

\section{\label{sec:NonlinearDelay}Forced DAO with nonlinear delayed feedback}

After studying the forced DAO a question arises: what are the differences between the forced DAO and the forced Duffing oscillator? In principle, both of them have the same nonlinear potential, a periodic forcing and enough dimensions to show chaos, and possibly Wada. However, there is an important feature that makes them different. 

In order to contrast the two systems, let us write on the one hand the forced Duffing oscillator equation as the following first order autonomous system,
\begin{align}
\begin{split}
\dot{x}_0 &= \omega \\
\dot{x}_1 &= -\gamma x_1 +x_2 +x_2^3+F \sin x_0 \\
\dot{x}_2 &= x_1,
\label{eq:DuffingExpanded}
\end{split}
\end{align}
On the other hand, following a usual technique for delay differential equations (see e.g. \cite{sprott_simple_2007}), we can rewrite the forced DAO of Eq. (\ref{eq:FalsoWada}) in form of an ODE with infinite dimensions: 
\begin{align}
\begin{split}
\dot{x}_0 &= \omega \\
\dot{x}_1 &= \alpha x_N +x_1 - (1+\alpha)x_1^3+F \sin x_0 \\
\dot{x}_i &= \dfrac{N}{\tau} (x_{i-1}-x_{i}), \textrm{ for } i\geq 2.
\label{eq:FalseWadaExpanded}
\end{split}
\end{align}
Comparing expressions (\ref{eq:DuffingExpanded}) and (\ref{eq:FalseWadaExpanded}) we see that they are very similar, but there is one important difference. In the case of the forced Duffing oscillator (Eq. \ref{eq:DuffingExpanded}), the evolution of $x_1$ depends on the nonlinear term of the variable $x_2$. However, in the forced DAO (Eq. \ref{eq:FalseWadaExpanded}) the evolution of $x_1$ depends linearly on $x_N$, and the nonlinearity is in $x_1$ (see table \ref{table:comparison} for an easy visualization of the two systems side by side).

\begin{table}[h!]
\centering
\begin{tabular}{|c|c|}
  \hline
Forced Duffing & Forced DAO expanded\\
  \hline
  \parbox{8cm}{
\begin{align*}
\begin{split}
\dot{x}_0 &= \omega \\
\dot{x}_1 &= -\gamma x_1 +x_2 +x_2^3+F \sin x_0 \\
\dot{x}_2 &= x_1,
\end{split}
\end{align*}} &
  \parbox{8cm}{
\begin{align*}
\begin{split}
\dot{x}_0 &= \omega \\
\dot{x}_1 &= \alpha x_N +x_1 - (1+\alpha)x_1^3+F \sin x_0 \\
\dot{x}_i &= \dfrac{N}{\tau} (x_{i-1}-x_{i}), \textrm{ for } i\geq 2.
\end{split}
\end{align*}}\\ 
  \hline
\end{tabular}
\caption{\label{table:comparison}Comparison between the forced Duffing and the forced DAO. The main difference between them is that the evolution of $x_1$ depends on the nonlinear term of $x_2$ in the forced Duffing, while it depends linearly on $x_N$ for the forced DAO.}
\end{table}

\begin{figure}
\begin{center}
\subfigure[]{\includegraphics[width=6cm]{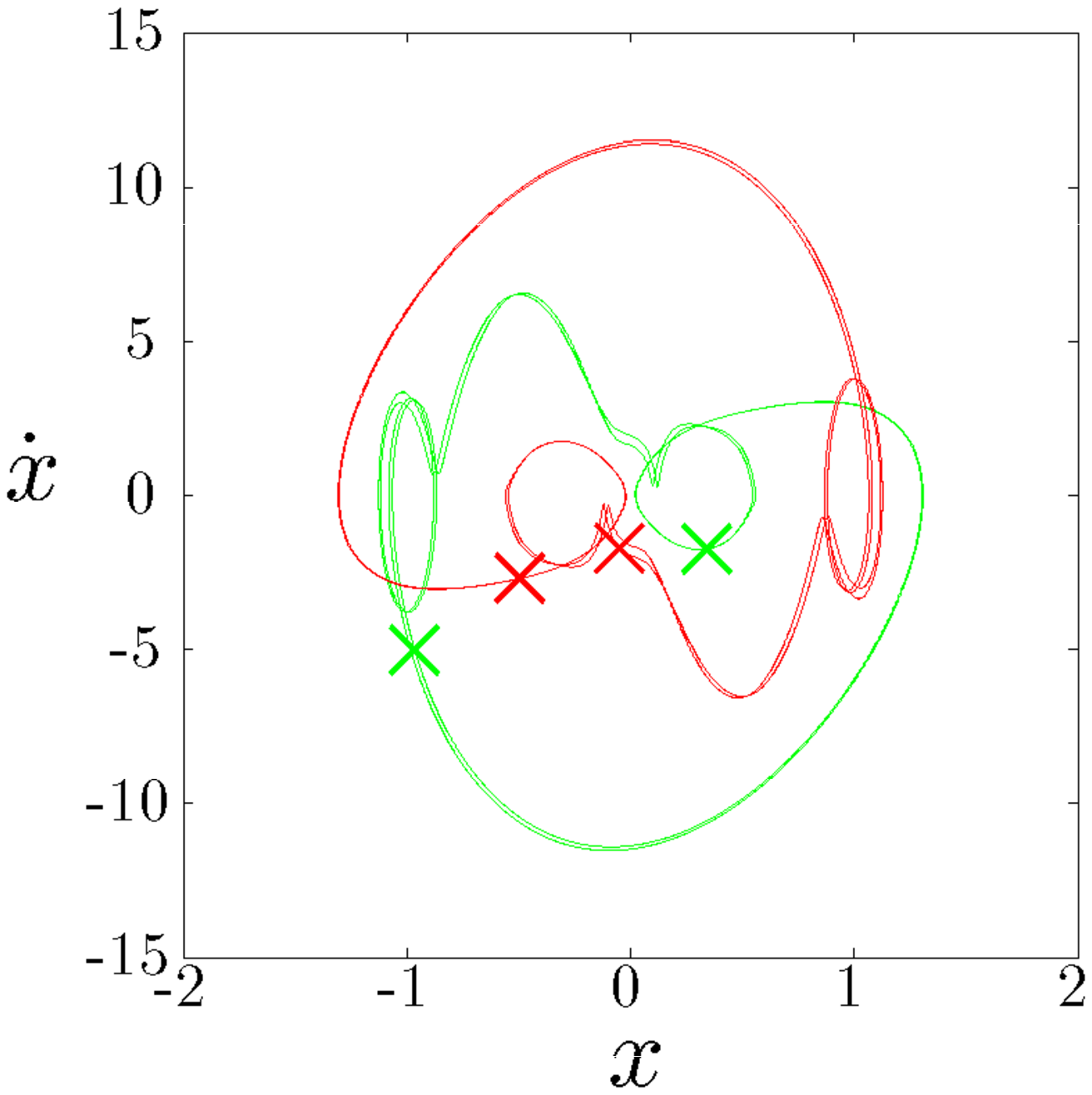}}
\subfigure[]{\includegraphics[width=6cm]{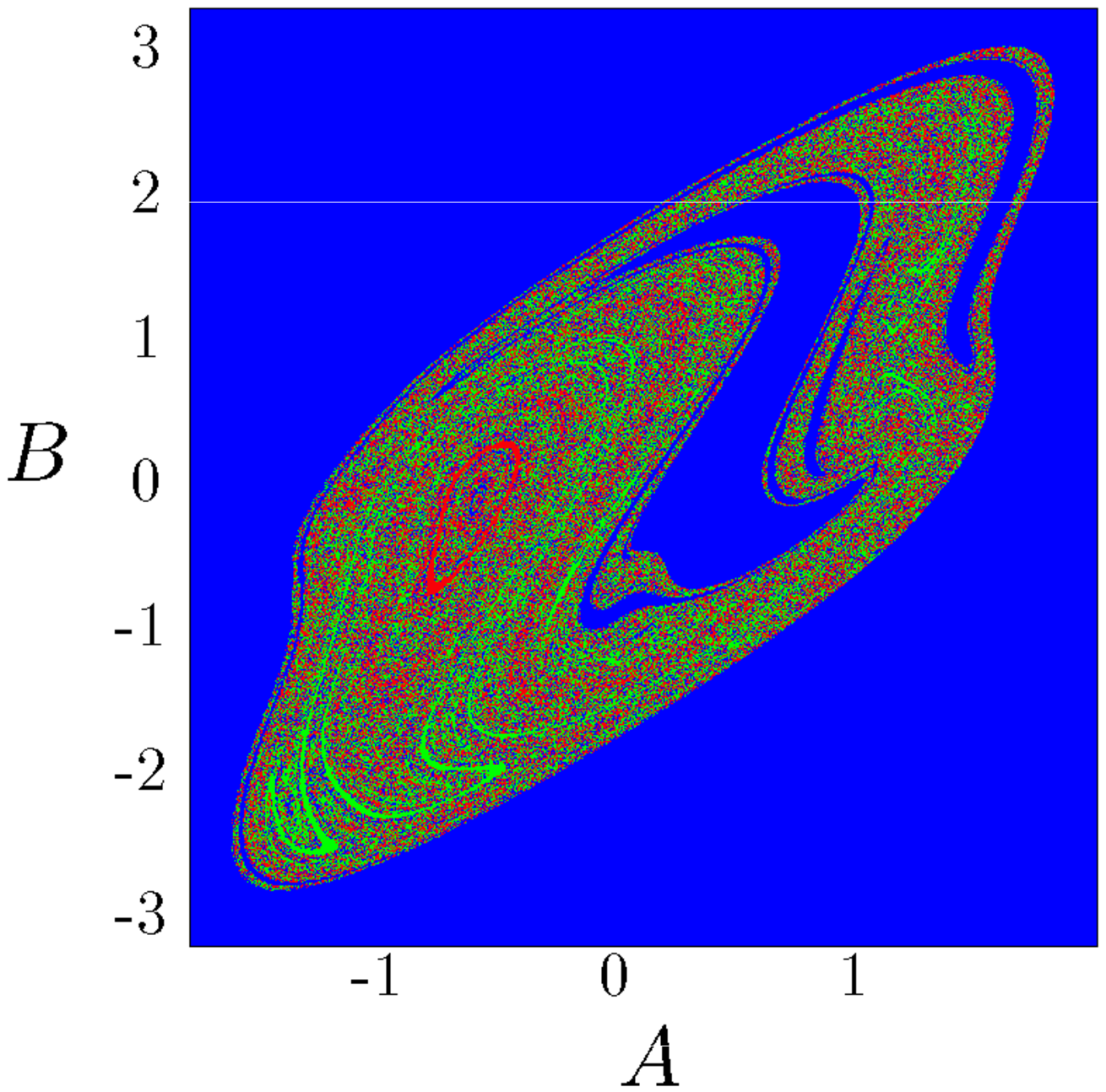}}
\subfigure[]{\includegraphics[width=6cm]{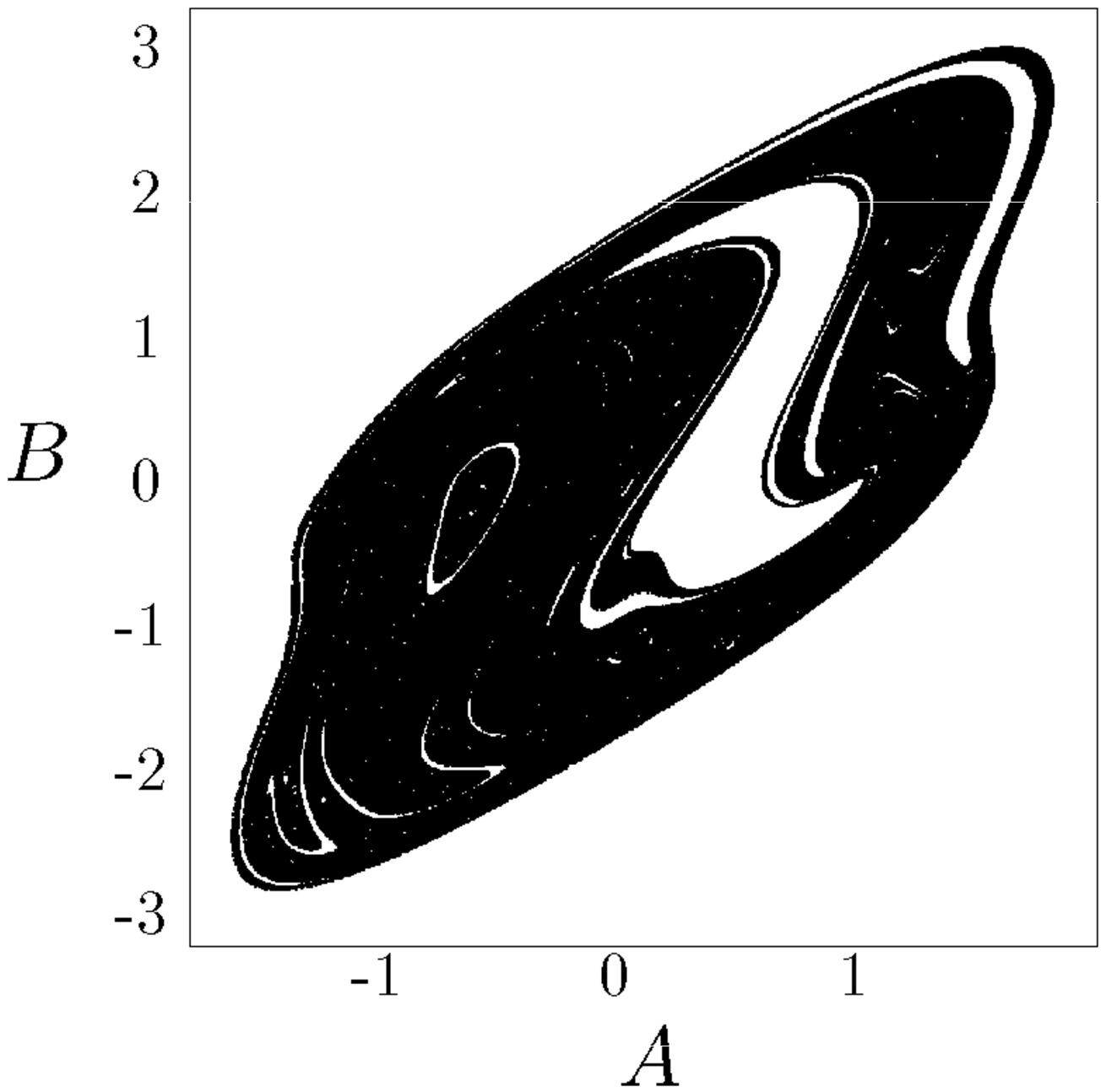}}
\subfigure[]{\includegraphics[width=6cm]{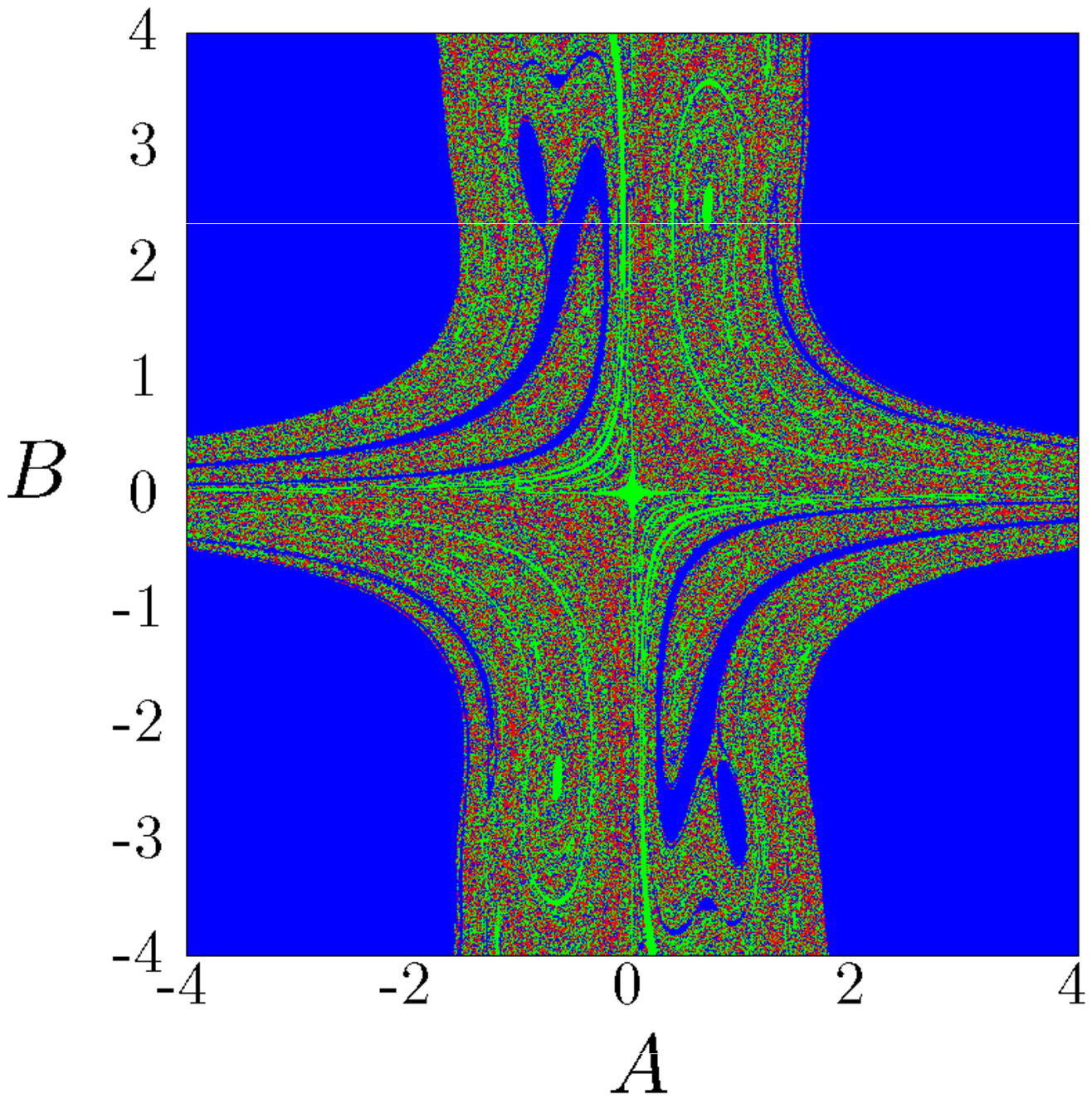}}
\end{center}
\caption{\label{fig:total} \textbf{Wada property induced by delay in nonlinear feedback}. (a) Attractors of the system $\dot{x}+ \alpha ( x_{\tau}^3 -x_{\tau}) +x=F \sin \omega t$ with  $\alpha=2.5$, $\tau=1$, $F=1.15$, and $\omega=1.2$. The system has two period-2 orbits and also diverging trajectories. Trajectories intersect themselves because this is a projection in $(x,\dot{x})$, but in principle, the system lives in infinite dimensions. (b) Basin of attraction: history functions leading to infinity are colored in blue, and the red and green colors are for history functions leading to the two period-2 orbits. This is an example of disconnected Wada basin. (c) All the points in the boundary separate three basins, thus the system possess the Wada property. (d) Basin of attraction with an oscillating history function $x(t)=A \sin Bt, \: \forall \, t \in \left[ -\tau,t_{0} \right]$. The Wada property is independent of the initial history function chosen ($W_3=1$). Different initial history functions are different subspaces of the same infinite dimensional space, they all have the Wada property.}
\end{figure}

As we mentioned earlier, the conditions for the Wada property are unknown, but in our exploration of delayed systems we find pertinent to study the system with the delay in the nonlinear term, looking even more similar to the forced Duffing oscillator. This system can be written in the usual manner as
\begin{equation}
\dot{x}+ \alpha ( x_{\tau}^3 -x_{\tau}) +x=F \sin \omega t. \label{eq:TotalWada}
\end{equation}
In the expanded version of this equation, we can see that the evolution of $x_1$ depends on the nonlinear term $x_N^3$. A careful exploration of the parameter space reveals that for $\alpha=2.5$, $\tau=1$, $F=1.15$, and $\omega=1.2$ the system has three attractors: one attractor at infinity (solutions that diverge) and the two period-2  attractors of Fig. \ref{fig:total}(a). Plotting the basin of attraction (Fig. \ref{fig:total}(b)),  we see that it looks like a disconnected Wada set. The method to verify the Wada property \cite{daza_testing_2015} confirms our intuition after a few steps: every point in the boundary separates three basins (see Fig. \ref{fig:total}(c)) giving a Wada parameter of $W_3=1$. The basins are disconnected because we are only looking at one slice of the infinite dimensional space of history functions, as it happens in the basins of the 3D scattering of Ref. \cite{kovacs_topological_2001}. For every history function we have tested, no matter how many parameters (dimensions) it had: the basin always shows the property of Wada. For example, Fig. \ref{fig:total}(c) is the plot of the basins of attraction for the same system with another family of history functions, a different slice of the infinite dimensional history function space, also showing the Wada property. These are solid arguments to affirm that the delay induces chaos and gives rise to the Wada property in the infinite dimensional space of history functions, turning the system strongly unpredictable and very sensitive to small changes in the history function.

\section{\label{sec:Discussion}Discussion}

In our exploration of the interplay between uncertainty and delay, we have investigated some simple delayed systems and their basins of attraction in the space of history functions. Given the apparent similarities between the delayed action and the Duffing oscillators, we have decided to add a periodic forcing to the delayed action oscillator and to look for the Wada property as it appeared in the basins of the Duffing oscillator. We have found the first example, to the best of our knowledge, where a linear delayed term induces transient chaos in a continuous system. Nonlinear delayed terms were known to induce chaotic dynamics \cite{mackey_oscillation_1977, heiden_dynamics_1982, hale_onset_1988}, but in this case the linear delay provides the extra dimensions that the system needs to show chaos. Although this constitutes an interesting result by itself, our objective is also to study the properties of the basins of attraction. Despite our careful research, we were unable to find the Wada property in the forced delayed action oscillator. Perhaps it happens in a small region of the parameter space, or perhaps it does not happen. Although the phase space is highly fractalized and the system is very close to show the full Wada property, we must label it as partially Wada. 

Finally, we introduced the delay in the cubic term. In this system we were able to find not only transient chaos, but also the full Wada property. The basins of attraction that we plot are subspaces of the infinite dimensional history function space, and all of them have the same properties. This means that this is probably the first report of the full Wada property in infinite dimensions. Infinite Wada basins can be obtained varying the family of history functions, and we can also modify the number of parameters obtaining Wada basins of arbitrary dimension. Without delay, this system would only show oscillatory dynamics, but here the delay induces both chaos and the Wada property. We expect that this study contributes to the investigation of delayed systems, especially concerning its sensitivity.

\section*{Acknowledgments}

This work was supported by Spanish Ministry of Economy and Competitiveness under Project No. FIS2013-40653-P.

\end{document}